\documentclass[11pt]{article} 
\usepackage{nips14submit_e,times}
\usepackage{hyperref}
\usepackage{url}

\usepackage[margin=1in]{geometry} 

\usepackage[ruled, vlined]{algorithm2e}

\usepackage{amsfonts}

\usepackage{amsmath}

\usepackage{graphicx}

\usepackage{multicol}

\usepackage[style=numeric, backend=bibtex]{biblatex}
\newcommand{\argmax}[1]{\underset{#1}{\mathrm{argmax}}}

\newcommand{\km}{$k$-mer}
\newcommand{\la}{\ensuremath{\leftarrow}}
\newcommand{\pa}{\textit{Pseudomonas aeruginosa}}
\newcommand{\phib}{\ensuremath{{\pmb \phi}}}

\newcommand{\rstar}{\ensuremath{{\mathcal{R}^\star}}}
\newcommand{\Scal}{\ensuremath{\mathcal{S}}}
\newcommand{\xb}{\ensuremath{\pmb x}}
\newcommand{\eqdef}{\overset{{\rm \mbox{\tiny def}}}{=}}

\title{Learning interpretable models of phenotypes from whole genome sequences with the Set Covering Machine}

\nipsfinalcopy 

\makeatletter
\newcommand*{\myfnsymbolsingle}[1]{%
  \ensuremath{%
    \ifcase#1
    \or 
      *%
    \or 
      \dagger
    \or 
      \ddagger
    \or 
      \mathsection
    \or 
      \mathparagraph
    \else 
      \@ctrerr  
    \fi
  }%
}   
\makeatother

\usepackage{alphalph}
\newalphalph{\myfnsymbolmult}[mult]{\myfnsymbolsingle}{}

\begin{document}

\maketitle

{\vspace{-20mm} \large
\begin{center}
Alexandre Drouin$^{1,}$\footnote{Corresponding author: \href{mailto:alexandre.drouin.8@ulaval.ca}{alexandre.drouin.8@ulaval.ca}\\{\indent Peer-reviewed and accepted for presentation at Machine Learning in Computational Biology 2014, Montr\'{e}al, Qu\'{e}bec, Canada.}}, S{\'e}bastien Gigu{\`e}re$^1$, Vladana Sagatovich$^1$, Maxime D{\'e}raspe$^2$,\\ Fran{\c c}ois Laviolette$^1$, Mario Marchand$^1$, Jacques Corbeil$^2$
\end{center}}

\begin{center}
\begin{tabular}{cc}
\multicolumn{2}{c}{\large Universit{\'e} Laval}\\
\multicolumn{2}{c}{\normalsize Qu\'{e}bec, Canada}\\
 & \\
${}^1$ D{\'e}partement d'informatique et de g{\'e}nie logiciel &${}^2$  D{\'e}partement de m{\'e}decine mol{\'e}culaire \\
\end{tabular}
\end{center}

\vspace{7mm}
\begin{abstract}

The increased affordability of whole genome sequencing has motivated its use for phenotypic studies. We address the problem of learning interpretable models for discrete phenotypes from whole genomes. We propose a general approach that relies on the Set Covering Machine and a \km~representation of the genomes. We show results for the problem of predicting the resistance of \pa, an important human pathogen, against 4 antibiotics. Our results demonstrate that extremely sparse models which are biologically relevant can be learnt using this approach.
\end{abstract}

\section{Introduction}

Recent advances in next-generation sequencing (NGS) have led to a tremendous increase in the affordability of whole genome sequencing (WGS)~\cite{vandijk2014}. 
The reduced cost and increased throughput of NGS have motivated the use of WGS for phenotypic study~\cite{vandijk2014,hall2013,koser2012,torok2012}.
Particularly, it is now possible to use whole genome sequences, instead of DNA microarrays, which require prior knowledge of the genomic regions of interest.
Learning models which can accurately predict a discrete phenotype from a genome has direct applications in the clinical setting and can lead to a better understanding of biological processes.
This is especially true if the learnt model is simple and easy to interpret, which, unfortunately, is not the case of most learning algorithms.
Moreover, the large size and increased availability of WGS data give rise to new computational challenges which must be addressed.

Machine learning algorithms require that the genomes be represented by a set of features. One common representation consists in a set of single nucleotide polymorphisms (SNP)~\cite{brookes1999}. 
Obtaining the SNPs for multiple genomes requires multiple sequence alignment, which is computationally expensive and affected by genome rearrangements~\cite{leimeister2014}. Moreover, potentially important information about the genome can be lost in this process.
In this paper we will favor an alternative approach, inspired from the ``bag-of-words'' model that is heavily used in the domain of text classification. It consists in representing each genome by the presence or absence of \km s, which are words of $k$ nucleotides that are possibly its subsequences~\cite{hall2013}. 
In addition, this approach does not require any sequence alignement~\cite{leimeister2014}.

As previously stated, another key requirement when learning to predict a phenotype is that the model must be interpretable by domain experts.
Interpretable models provide biological insight on the decision function and can lead to the discovery of novel biological processes.
Models must thus be sparse and composed of elements from which sufficient biological knowledge can be extracted. 
Sparsity of the model is desirable since it contributes to reducing the cost of validation and promotes its usage in a clinical setting. 
Many state-of-the-art learning algorithms do not provide interpretable models. 
This is the case of the Support Vector Machine (SVM)~\cite{cortes1995}, which yields dense models. 
The Lasso~\cite{tibshirani1996} yields sparse models compared to the SVM.
Nevertheless, they often contain a large number of non-null coefficients, rendering their biological understanding difficult.

In this paper, we propose a novel approach for learning sparse and interpretable models from whole genomes for predicting discrete phenotypes.
Our approach relies on the Set Covering Machine~\cite{marchand2003}, a learning algorithm that produces highly sparse models that achieved state-of-the-art accuracy for many learning task, such as learning from DNA microarray data~\cite{shah2012}. 
The models obtained are short conjunctions or disjunctions of boolean valued rules which can explicitly highlight the importance of specific DNA sequences.

We first present the Set Covering Machine learning algorithm. 
Subsequently, we demonstrate how it can be applied to predict phenotypes based on whole genomes. 
We then present an example application to the problem of predicting the antimicrobial resistance of \pa~(PA), an opportunistic, nosocomial pathogen of immunocompromised individuals~\cite{kos2014}.
PA typically infects the pulmonary tract, urinary tract, burns, wounds, and also causes other blood infections. 
It is an important human pathogen.
Finally, we discuss the models found for the resistance of PA against four antibiotics, the limitations of the approach, and propose paths to extend our work.

\section{Methods}

\subsection{The Set Covering Machine}

In the supervised machine learning setting, we assume that data are available as a set $\Scal=\{(\xb_i,y_i)\}_{i=1}^m \sim D^m$, where $\xb_i \in \mathcal{X}$ is a training example, $y_i \in \mathcal{Y}$ its associated label and $D$ is a data generating distribution. 
We consider binary classification problems where $\mathcal{Y} = \{0, 1\}$. 
The goal of a learning algorithm is to obtain a predictor $h: \mathcal{X} \rightarrow \mathcal{Y}$, such that $h(\xb) = y$ for most $(\xb, y) \sim D$. 
The Set Covering Machine (SCM)~\cite{marchand2003} is a learning algorithm that produces predictors that are conjunctions or disjunctions of boolean valued rules $r_i : \mathcal{X} \rightarrow \{0, 1\}$. 
Given a set of rules, the SCM attempts to find the smallest subset of rules that correctly classifies the data. 
As mentioned in~\cite{marchand2003}, this problem can be reduced to the set cover problem which is known to be NP-hard.
However, the SCM uses a greedy algorithm to obtain an approximate solution with a worst case guarantee. 
Algorithm \ref{algo:train_scm} presents the SCM algorithm in the case where the returned predictor is a conjunction of boolean valued rules. 
The disjunction case can be obtained from the previous one by using $\Scal'=\{(\xb_i, \neg y_i) : (\xb_i, y_i) \in \Scal\}$ as the set of training examples for Algorithm \ref{algo:train_scm} and taking the complement of the returned predictor $h$. 
This follows from the De Morgan law: $\neg \bigwedge_{r^\star_i \in \rstar} r^\star_i(\xb) = \bigvee_{r^\star_i \in \rstar} \neg r^\star_i(\xb) $.

In Algorithm~\ref{algo:train_scm}, the parameter $s$ acts as a regularizer by limiting the complexity of $h$. 
The parameter $p$ adds robustness to noise and to class imbalance, by controlling a trade-off between minimizing the error on each class. 
If more than one rule have a maximal utility value $U_i$, we select the one which most reduces the empirical risk and thus maximises $|\mathcal{A}_i| - |\mathcal{B}_i|$.
If there still exists a tie, we select either of the rules.
Notice that, at each iteration, only the examples for which the outcome of $h$ is not already determined are considered. 
This prevents the selection of correlated rules, which would unnecessarily increase the complexity of the predictor. 
This is crucial in a clinical setting, since increased predictor complexity can lead to dispensable costs. 

The running time complexity of Algorithm \ref{algo:train_scm} is $O(|\mathcal{R}| \cdot |\Scal| \cdot s)$ in the worst case. 
It thus scales linearly in the number of rules and the number of training examples.
In addition, note that at each iteration, the computation can be parallelized by distributing the computation of the $|\mathcal{A}_i|$ and $|\mathcal{B}_i|$ on multiple cores.

\begin{algorithm}
\DontPrintSemicolon
\SetAlgoLined
\SetKwInOut{Input}{input}  
\Input{$S$: A set of training examples,
       $\mathcal{R}$: A set of boolean valued rules,
       $p$: The class tradeoff parameter,
       $s$: The maximum number of rules in $h$.}

 $\rstar \la \emptyset$ \;
 $\mathcal{P} \la \mbox{the set of examples in $\Scal$ with label $1$}$ \;
 $\mathcal{N} \la \mbox{the set of examples in $\Scal$ with label $0$}$ 
 
 $stop \la False$  
 
 \While{$\mathcal{N} \not= \emptyset$ {\upshape \textbf{and}} $|\rstar| < s$ {\upshape \textbf{and}} $\neg stop$}{
   $\forall r_i \in \mathcal{R},~\mathcal{A}_i \la$ the subset of $\mathcal{N}$ correctly classified by $r_i$ \;
   $\forall r_i \in \mathcal{R},~\mathcal{B}_i \la$ the subset of $\mathcal{P}$ misclassified by $r_i$ \;

   $\forall r_i \in \mathcal{R},~U_i \la |\mathcal{A}_i| - p \cdot |\mathcal{B}_i|$ \textbf{if} $|\mathcal{A}_i| \geq |\mathcal{B}_i|$ \textbf{and} $-\infty$ \textbf{otherwise}\;
   $i^\star \la \argmax{i}~U_{i}$ \;
   \eIf{$U_{i^\star} \not= -\infty$}
   {
     $\rstar \la \rstar \cup \{r_{i^\star}\}$ \;
     $\mathcal{N} \la \mathcal{N} - \mathcal{A}_{i^\star}$ \;
     $\mathcal{P} \la \mathcal{P} - \mathcal{B}_{i^\star}$ \;
   }
   {$stop = True$}
 }
 
 \Return $h$, where $h(\xb) = \bigwedge_{r^\star_i \in \rstar} r^\star_i(\xb)$

 \caption{Train SCM($S$, $\mathcal{R}$, $p$, $s$)}
 \label{algo:train_scm}
\end{algorithm}

\subsection{Applying the Set Covering Machine to whole genomes}

In order to apply the SCM to whole genomes, we use a \km~representation. 
First, given a dataset $\Scal=\{(\xb_i,y_i)\}_{i=1}^m$, where $\xb_i$ is a genome and $y_i \in \{0, 1\}$ is the phenotype, we define $\mathcal{K}$ as the set of all unique \km s that are at least in one genome. 
All overlapping \km s are considered while constructing this set.
Note that \km s that are in all genomes can be discarded, since they cannot be selected by Algorithm \ref{algo:train_scm}.
Then, for each $(\xb, y) \in \Scal$, we represent a genome $\xb$ by $\phib(\xb)$, a boolean vector of $|\mathcal{K}|$ dimensions where component $\phi_i(\xb) = 1$ if $k_i \in \mathcal{K}$ is in genome $\xb$ and $0$ otherwise.
Then, for each \km~$k_i \in \mathcal{K}$, we  define a \emph{presence rule}\, $p_{k_i}(\phib(\xb)) \eqdef I(\phi_i(\xb) \!=\! 1)$ and an \emph{absence rule}\, $a_{k_i}(\phib(\xb))  \eqdef I(\phi_i(\xb) \!=\! 0)$, where $I$ is the indicator function.
We can then apply Algorithm \ref{algo:train_scm} by taking $\Scal'=\{(\phib(\xb_i), y_i) : (\xb_i, y_i) \in \Scal\}$  as the training set of examples and by using the set of $2 \cdot |\mathcal{K}|$ presence and absence rules for $\mathcal{R}$. 
By doing so, we obtain a predictor which explicitly highlights the importance of a small set of \km s for predicting a phenotype. 
This predictor has a form which is simple to interpret, since its predictions are the outcome of a simple logical operation. The running time complexity of using Algorithm \ref{algo:train_scm} in this setting is $O(|\mathcal{K}| \cdot |\Scal| \cdot s)$.

\subsection{Predicting the antimicrobial resistance of \pa}
We validated our approach by addressing the 
problem of antibiotic resistance.
There is a pressing need for rapid clinical diagnostic tests, that can accurately predict the resistance of a bacterial strain to a wide range of antibiotics~\cite{tuite2014}.
Our approach could be used to obtain interpretable predictors of the resistant phenotype for multiple antibiotics.
In addition to assist in a clinical setting, these may help to uncover new resistance mechanisms and ultimately identify new drug targets. 
In order to learn such models, we used a dataset containing the genomes of 390 \pa~strains and their measured resistance phenotype (resistant, intermediate or sensitive) for 4 antibiotics~\cite{kos2014}.
We addressed each antibiotic as a distinct classification problem.
We binarized each problem by discarding intermediate strains and assigning the $1$ label to resistant strains and the $0$ label to sensitive strains.
For constructing the \km~set~$\mathcal{K}$, we used $k=31$, because it is a standard parameter used in \emph{de novo} genome assembly algorithms~\cite{boisvert2010ray}; if  $k=31$ provides enough information to allow good genome assembly, it should be suitable for our learning task.
An overview of the resulting dataset is shown in Table \ref{tbl:pa_data}.

For each antibiotic, we obtained sets $\mathcal{K}$ which contained millions of \km s. 
Therefore, the dimensionality of $\phib(\xb)$ was extremely high, leading to the problem of storing the examples in memory. 
In order to overcome this problem, we stored the $\phib(\xb)$ vectors on disk in an HDF5 dataset~\cite{hdf5}, which supports built-in compression and array-like access to the data. 
At each iteration of Algorithm \ref{algo:train_scm}, we accessed the data by blocks of a fixed number of rows and columns. 
This led to an implementation for which the memory usage was fully controllable. 
For each antibiotic, the first iteration of Algorithm \ref{algo:train_scm} took less than 7 minutes, while using a single CPU and 3 GB of RAM on a system equipped with a 2.8 GHz Intel Core i7 and a 5200 RPM hard drive. All subsequent iterations took less time.

\begin{table}
\centering
\begin{tabular}{|l||c|c|c|c|}
Antibiotic & Resistant & Sensitive & Total & $2 \cdot |\mathcal{K}|$\\ \hline
Amikacin & 84 & 281 & 365 & 119 023 612\\
Doripenem & 137 & 226 & 363 & 118 580 512\\
Levofloxacin & 169  & 189 & 358 & 118 335 666\\
Meropenem & 153 & 215 & 368 & 118 931 911\\
\end{tabular}
\caption{Number of resistant/sensitive strains and boolean valued rules ($2 \cdot |\mathcal{K}|$) for each antibiotic}
\label{tbl:pa_data}
\vspace{-0.5 cm}
\end{table}

\section{Results and discussion}

\subsection{Evaluating the performance of the proposed approach}
For each antibiotic, we conducted nested 5-fold cross-validation (CV). 
We first split the dataset into 5 parts, called the \emph{outer-folds}. 
One outer-fold was left out as a validation set to compute the risk of the algorithm, while the remaining outer-folds formed a training dataset used to train the algorithm. On this \emph{inner dataset}, we performed standard 5-fold CV to select the algorithm's hyperparameters. After choosing those minimizing the \emph{inner CV} risk, we retrained the algorithm using the whole inner dataset, and computed predictions for the examples of the left out outer-fold. We repeated this procedure 5 times (once per outer fold) and reported the mean left out outer-fold risk, which corresponds to an estimate of the risk incurred on novel data.

We conducted this experiment for our approach and selected hyperparameters~$p$ and~$s$ of Algorithm \ref{algo:train_scm} from ranges $[10^{-1}, 10^1]$ and $\{1, ... , 10\}$ respectively.
For comparison, we repeated the experiment using a SVM with a linear kernel and selected the values of hyperparameter $C$ in range $[10^{-5}, 10^9]$. \footnote{Note that learning with the SVM from the $\phib(x)$ vectors is equivalent to using the spectrum kernel~\cite{leslie2002}.}
In addition, in order to validate that our approach actually learns from the WGS data, we compared to a predictor that predicts the majority class for each antibiotic.
The results are summarized in Table \ref{tbl:nested_5_5_benchmark}.
On all but one antibiotic, our approach outperforms the SVM.
This is an interesting result, since our predictors are composed of very few rules, whereas the SVM decision function is fully dense and thus attributes non-null weights to millions of rules.
Moreover, these results confirm that our approach indeed learns from the WGS data, since it outperforms the majority class predictor.

\begin{table}
\centering
\begin{tabular}{|c||c|c||c|c|}
Antibiotic & \multicolumn{2}{|c||}{SCM} & SVM & Majority\\
~ & Risk & $|\rstar|$ & Risk & Risk\\ \hline
Amikacin & \textbf{0.170} $\pm$ 0.024 & 3.0 $\pm$ 0.9 & 0.189 $\pm$ 0.032 & 0.230 $\pm$ 0.036\\
Doripenem & 0.283 $\pm$ 0.046 & 1.4 $\pm$ 0.5 &  \textbf{0.270} $\pm$ 0.012 & 0.378 $\pm$ 0.040\\
Levofloxacin & \textbf{0.075} $\pm$ 0.025 & 1.6 $\pm$ 0.5 & 0.232 $\pm$ 0.029 & 0.472 $\pm$ 0.034\\
Meropenem & \textbf{0.288} $\pm$ 0.018 & 1.0 $\pm$ 0.0 & 0.340 $\pm$ 0.043 & 0.416 $\pm$ 0.048\\
\end{tabular}
\caption{All values presented in this table are means over the 5 outer-folds followed by their standard deviation. Hence, risks for the SCM, the SVM and the majority class predictor are given for each antibiotic. For the SCM, we also show the  number of rules $|\rstar|$ used. The best risk values are in bold. }
\label{tbl:nested_5_5_benchmark}
\end{table}

\subsection{Biological relevance of the predictors} 
To evaluate the ability of our approach to learn biologically relevant models, we relearned an SCM on  \underline{the entire} dataset of each antibiotic.
The optimal parameters for each dataset were selected by 5-fold CV based on the same ranges as in the previous experiment. 
For amikacin and meropenem, the obtained predictors are conjunctions of 5 and 2 rules respectively.
For doripenem and levofloxacin, the predictors are disjunctions of 2 rules. Hence, in each case, the obtained predictor is based on a very small number of rules.

The predictor for levofloxacin is a disjunction of 2 absence rules for \km s located in the wildtype DNA gyrase, which is the target of levofloxacin.
Interestingly, the first \km~is located in the quinolone-resistance-determining region of subunit A and covers two amino acids, Thr-83 and Asp-87, known to confer resistance to levofloxacin when mutated~\cite{akasaka2001}. 
Similarly, the second \km~is located in subunit B and covers amino acids Ser-468 and Glu-470, which also confer resistance when mutated~\cite{akasaka2001}. Therefore, in this case, the predictor has recovered known biological facts.
For all other antibiotics, the model contains a rule relative to the absence of a \km~in the DNA gyrase. However, none of these antibiotics actually target the DNA gyrase. 

\begin{figure}[h]
\begin{center}
\includegraphics[width=0.42\linewidth]{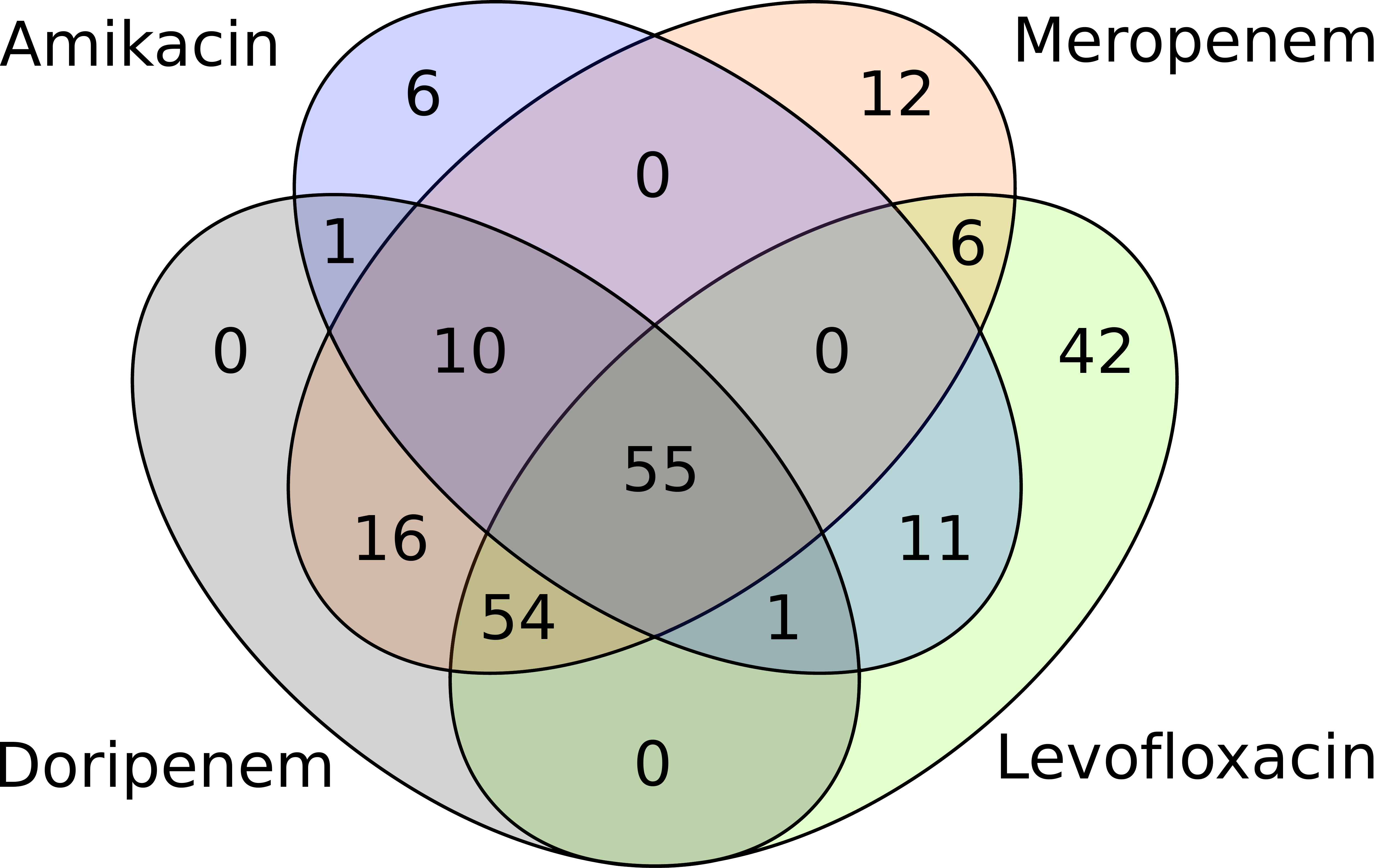}
\end{center}
\vspace{-0.5cm}
\caption{Distribution of the resistant strains among the antibiotics in the dataset.}
\label{fig:resistance_overlap}
\end{figure}

In Figure \ref{fig:resistance_overlap}, we see that most strains are resistant to multiple antibiotics and that many are resistant to all of them (central set). 
Following this observation, we used each predictor to predict the labels of the training examples.
We analysed the distribution of false negatives among the sets of Figure~\ref{fig:resistance_overlap} and found that only a slight fraction fall in the central set.
This suggests that, since the SCM can only learn one conjunction or disjunction, it learns the one which best classifies the largest set of the training examples.
The presence of the DNA gyrase in all the predictors can be explained by a shared resistance with levofloxacin for many examples.
In addition, this observation supports the difference in the accuracies observed for levofloxacin and all other antibiotics in Table~\ref{tbl:nested_5_5_benchmark}.

\section{Conclusion}

We have addressed the problem of learning interpretable models of phenotypes from whole genome sequences. 
We have demonstrated that the Set Covering Machine can be used to achieve this goal. 
Our results for the problem of predicting the antibiotic resistance of \pa~suggest that our approach indeed yields sparse and interpretable models. 
For one antibiotic, we have recovered the target gene and for the others, the models were interpretable enough to gain insight on a limitation of our approach.
Future works will therefore address the problem of learning more than one conjunction/disjunction with the SCM. A disjunction of conjunctions can still be very sparse, and may allow to model richer biological pathways than a single conjunction or a single disjunction.

\subsubsection*{Acknowledgments}

The authors would like to thank Veronica Kos, Humphrey Gardner and their colleagues from AstraZeneca for providing the \pa~dataset. Computations were performed on the Colosse supercomputer at Universit\'{e} Laval (resource allocation project: nne-790-ad), under the auspices of Calcul Qu\'{e}bec and Compute Canada. AD is recipient of an Alexander Graham Bell Canada Graduate Scholarship Doctoral Award from the National Sciences and Engineering Research Council of Canada (NSERC). This work was supported in part by the Fonds de recherche du Qu\'{e}bec - Nature et technologies (FL, MM \& JC; 2013-PR-166708) and the NSERC Discovery Grants (FL; 262067, MM; 122405). JC acknowledges the Canada Research Chair in Medical Genomics.

\printbibliography

\end{document}